\documentclass[11pt,a4paper]{article}
\usepackage[a4paper,total={160mm,220mm}]{geometry}
\usepackage{amsmath,amssymb}
\usepackage{cite}
\usepackage{braket}
\usepackage{hyperref}
\usepackage[capitalise]{cleveref}
\usepackage{caption}
\usepackage{subcaption}
\usepackage{graphicx}
\graphicspath{{figures/}}

\newcommand{\abs}[1]{\left\lvert{#1}\right\rvert}
\newcommand{\absl}[1]{\lvert{#1}\rvert}
\newcommand{\md}{\mathrm{d}}
\newcommand{\me}{\mathrm{e}}

\renewcommand{\vec}[1]{\mathbf{#1}}

\begin{document}

\numberwithin{equation}{section}
\title{
\vspace*{-0.5cm}
{\scriptsize \mbox{}\hfill MITP-25-047}\\
\vspace{3.5cm}
\Large{\textbf{Exploring Nonperturbative Behaviour of\\Moments and Cumulants in Quantum Theories}}
\vspace{0.5cm}
}

\author{Sebastian Schenk\\[2ex]
\small{\em PRISMA$^+$ Cluster of Excellence \& Mainz Institute for Theoretical Physics,} \\
\small{\em Johannes Gutenberg-Universit\"at Mainz, 55099 Mainz, Germany}\\[0.8ex]}

\date{}
\maketitle

\begin{abstract}
\noindent
The dynamics of quantum fields become nonperturbative when their interactions are probed by a large number of particles.
To explore this regime we study correlation functions which involve a large number of fields, focussing on massive scalar theories that feature arbitrary self-interactions, $\phi^{2p}$.
Treating quantum fields as operator-valued distributions, we investigate $n$-point correlation functions at ultra-short distances and compute moments and cumulants of fields, using a semiclassical saddle point approximation in the double scaling limit of weak coupling, $\lambda \to 0$, large quantum number, $n \to \infty$, while keeping $\lambda n$ constant.
Addressing the nonperturbative regime, where $\lambda n \gtrsim 1$, requires a resummation of the effective saddle point to all orders in $\lambda n$.
We perform this resummation in zero and one dimensions, and show that the moments, corresponding to correlation functions including disconnected contributions, grow exponentially with $n$.
This growth is significantly reduced for higher-order self-interactions, i.e.~for larger $p$.
On the other hand, we argue that the cumulants, which represent connected correlation functions, grow even more rapidly and are mostly independent of $p$.
\end{abstract}

\newpage

\section{Introduction}
\label{sec:introduction}

Modern quantum field theory (QFT) calculations achieve a remarkable precision that is consistently confirmed by experimental results.
This high level of accuracy is facilitated by the powerful framework of perturbation theory, which employs formal series expansions to systematically organise small deviations from the vacuum state at weak coupling.
However, a comprehensive assessment of QFT dynamics far away from the vacuum is still lacking.
These dynamics typically include particle interactions at extremely high energies, for instance through scattering of many quanta, thereby probing the ultraviolet (UV) properties of fields.

Despite its power, perturbation theory has been observed to break down at very high energies across various scenarios, including the electroweak instanton sector~\cite{Ringwald:1989ee, Espinosa:1989qn, McLerran:1989ab}, generic scalar QFTs~\cite{Cornwall:1990hh, Goldberg:1990qk, Brown:1992ay, Voloshin:1992mz, Argyres:1992np, Voloshin:1992nu, Smith:1992rq, Libanov:1994ug, Khoze:2014zha, Son:1995wz, Khoze:2017ifq, Khoze:2018kkz, Schenk:2021yea}, or massless string scattering amplitudes~\cite{Ghosh:2016fvm, Ghosh:2017pel}.
This breakdown is typically attributed to the divergent, asymptotic nature of perturbative expansions~\cite{Dyson:1952tj}, and associated with the rapidly growing number of Feynman diagrams at each order of the series~\cite{Hurst:1952zh, Bender:1976ni}.
Intuitively, one would expect this growth, and hence the corresponding energy scale at which nonperturbative effects emerge to depend on the underlying QFT model.
For instance, in multiparticle scattering processes, higher-order self-interactions, $\phi^{2p}$ for integer $p$, in principle generate an arbitrary number of particles at each interaction vertex, suggesting that such interactions could significantly enhance particle production rates.
On the other hand, the staggering number of Feynman diagrams and their interference effectively can lead to low-point interactions, such as $\phi^3$ or $\phi^4$, governing $n$-particle rates at the kinematic threshold~\cite{Khoze:2017lft}.
This results in an intricate interplay between different operators in an effective field theory~\cite{Khoze:2022fbf}.

In this work, we shed some more light on this question by addressing a conceptually identical yet technically simpler aspect of the problem.
Specifically, we investigate $n$-point correlation functions, which serve as a prototype for field dynamics where nonperturbative behaviour is expected as $n$ increases.
These scenarios are typically analysed through a semiclassical expansion around a nontrivial saddle point that accounts for the large number of field insertions, often enabling analytic~\cite{Khlebnikov:1992af, Son:1995wz, Khoze:2017ifq, Khoze:2018kkz, Khoze:2018mey} as well as numerical treatments~\cite{Demidov:2021rjp, Demidov:2022ljh, Demidov:2023dim}.
For example, the scaling dimension of the operator $\phi^n$ in conformal field theories (CFTs) at the Wilson-Fisher fixed point has been determined in the double scaling limit of perturbatively small fixed-point coupling, $\lambda \to 0$, large quantum number, $n \to \infty$, while keeping $\lambda n$ fixed~\cite{Badel:2019oxl, Badel:2019khk, Antipin:2024ekk}, and recently violations of superadditivity in large-charge sectors of CFTs have been explored~\cite{Cohen:2025skq}.

Here, we take initial steps to go beyond CFTs by exploring the behaviour of large-$n$ correlation functions of massive fields in the short-distance limit, where all fields are inserted at the same spacetime point, $\braket{\phi^n(x)}$.
While these expectation values of composite operators can be formally understood as the limit of correlation functions, where $\absl{x_i - x_j} \to 0$ for all spacetime points, they diverge in this regime, revealing the UV behaviour of the QFT.
Consequently, composite operators require a careful renormalisation, for instance provided by the operator product expansion~\cite{Hollands:2023txn}.
To regulate the divergent behaviour, we adopt an axiomatic formulation of QFT, treating quantum fields as operator-valued distributions with test functions of localised support~\cite{Wightman:1956zz, StreaterWightman2001, Jaffe:1967nb}.
Physically, the local spacetime points where the fields are inserted cannot be distinguished beyond a finite energy scale.
In this context, the correlation functions correspond to moments and cumulants of the probability density defined by the classical action.
To compute these quantities we first employ a semiclassical approach expanding around a nontrivial saddle point of an effective action that explicitly accounts for the large number of fields, in the double scaling limit $\lambda \to 0$ and $n \to \infty$ with $\lambda n$ fixed.
However, this saddle point expansion is only accurate in the perturbative regime where $\lambda n \ll 1$.
In zero and one dimensions, we perform a resummation of this expansion to all orders in $\lambda n$, allowing us to investigate the nonperturbative regime where $\lambda n$ is no longer small, $\lambda n \gtrsim 1$.
In this domain, the moments, which correspond to correlation functions that include disconnected contributions, grow exponentially.
This growth is significantly reduced for higher-order self-interactions, i.e.~for larger values of $p$.
In contrast, the cumulants, representing fully-connected correlation functions that encapsulate all physical information of the quantum theory, exhibit universal growth with $n$.
That is, their growth is mostly independent of $p$.
While this result provides insights into the nonperturbative behaviour of large-$n$ correlation functions in quantum theories, we should nevertheless be aware that realistic QFT settings introduce additional complications, such as a nontrivial phase space or the renormalisation of operators.

This work is organised as follows.
In \cref{sec:CorrelationFunctions}, we briefly review connected as well as nonconnected correlation functions in QFTs from a path integral perspective.
In \cref{sec:OrdinaryIntegrals}, we build intuition for these correlation functions by examining ordinary integrals as an analogue of a QFT in zero dimensions.
We then scrutinise this intuition for functional path integrals in \cref{sec:QMandQFT}, where we employ a semiclassical approach to gain insights into the large-$n$ behaviour of correlation functions.
Finally, we summarise our findings and conclude in \cref{sec:conclusions}.

\section{Correlation Functions Involving a Large Number of Fields}
\label{sec:CorrelationFunctions}

In a quantum theory, all physical information such as the spectra of excitations and their scattering amplitudes is encoded in fully connected correlation functions of local operators.
If one were to know all correlation functions precisely, one would effectively have ``solved" the quantum theory.
In practice, these correlation functions are often determined by evaluating Feynman diagrams with fully connected topologies.
This construction corresponds to a perturbative expansion in terms of a small parameter, typically the coupling times Planck's constant, $\lambda \hbar$.

\subsection{Generating functionals for correlation functions}

Formally, connected correlation functions can be derived from their nonconnected counterparts, i.e.~correlation functions that also include disconnected contributions such as vacuum bubbles.
These are defined as vacuum expectation values of time-ordered operator products,
\begin{equation}
	\gamma_n (x_1, \ldots, x_n) = \braket{0 | \phi(x_1) \ldots \phi(x_n) | 0} \, .
\end{equation}
They are generated by the partition function $Z[J]$ of the quantum theory, such that
\begin{equation}
	\gamma_n (x_1, \ldots, x_n) = \left. \frac{\delta^n}{\delta J(x_1) \ldots \delta J(x_n)} Z[J] \right\rvert_{J=0} \, .
	\label{eq:NonconnectedGnGeneratingFunction}
\end{equation}
For a theory with Euclidean action $S_E$, the partition function is defined as
\begin{equation}
	Z[J] = \int \mathcal{D} \phi \, \me^{-S_E + \int \md^d x \, J(x) \phi(x)} \, .
\end{equation}
The partition function measures the response of the field dynamics to an external local source $J(x)$, which enters the equation of motion as an inhomogeneous source term.
In other words, it quantifies the probability of the vacuum state in the asymptotic past remaining in the vacuum state in the asymptotic future, $Z[J] = \braket{0 | 0}_J$, under the influence of a driving force $J$.

The connected correlation functions correspond to the cumulants constructed from the nonconnected correlation functions.
For instance, the one-point function in the presence of the external source satisfies the relation
\begin{equation}
	G^J_1 (x) = \frac{\gamma^J_1 (x)}{Z[J]} \, .
\end{equation}
Here, the superscript $J$ indicates that the source has not yet been set to zero.
Differentiating this expression with respect to the source and subsequently setting $J=0$ yields the cumulants of arbitrary order, for instance
\begin{equation}
	G_1(x) = \frac{\gamma_1(x)}{Z_0} \, , \enspace
	G_2(x,y) = \frac{\gamma_2(x,y)}{Z_0} - \frac{\gamma_1(x) \gamma_1(y)}{Z_0^2} \, ,
\end{equation}
where $Z_0 = Z[0]$.
In general, following this prescription, the connected correlation functions are generated by the logarithm of the partition function,
\begin{equation}
	G_n (x_1, \ldots, x_n) = \left. \frac{\delta^n}{\delta J(x_1) \ldots \delta J(x_n)} \ln Z[J] \right\rvert_{J=0} \, .
	\label{eq:ConnectedGnGeneratingFunction}
\end{equation}
In terms of probability theory, the correlation functions $\gamma_n$ and $G_n$ correspond to the moments and cumulants of the probability density defined by $\exp(-S_E)$, respectively.

For this work's purpose, we investigate the nonperturbative behaviour of correlation functions at high energies, which is governed by both the number of field insertions $n$ and their short distance limits.
In this regime, the spacetime points where the field operators are inserted become indistinguishable below a certain energy scale.
Consequently, the objects of interest include quantities such as $\braket{\phi^2(x)}$ or $G_2(x,x)$.
When evaluating connected correlation functions at coincident spacetime points, we can calculate \cref{eq:ConnectedGnGeneratingFunction} explicitly using Fa\`a di Bruno's formula (see, e.g., \cite{Fraenkel1978Formulae}), yielding
\begin{equation}
	G_n (x, \ldots , x) = \sum_{k=1}^n (-1)^{k-1} (k-1)! B_{n,k} \left(\frac{\gamma_1}{Z_0} , \ldots , \frac{\gamma_{n-k+1}}{Z_0} \right) \, .
	\label{eq:ConnectedGnFaaDiBruno}
\end{equation}
Here, $B_{n,k}$ denote the partial exponential Bell polynomials and the nonconnected correlation functions correspond to vacuum expectation values of the form $\gamma_n = \braket{0 | \phi^n(x) | 0}$.
Due to Poincar\'e symmetry, the choice of $x$ is arbitrary.

However, it is important to note that the formal expressions in \cref{eq:NonconnectedGnGeneratingFunction,eq:ConnectedGnGeneratingFunction} are valid only for local field operators which are inserted at distinct spacetime points, $x_i \neq x_j$.
As their separation approaches zero, $\absl{x_i - x_j} \to 0$, we expect these expressions to diverge unless the UV behaviour of the quantum theory is properly regulated.
For instance, the propagator of a free real scalar field behaves as 
\begin{equation}
	\braket{0 | \phi(x) \phi(y) |0} \sim \frac{1}{\abs{x-y}^2} \, ,
\end{equation}
which diverges at short distances as $\absl{x-y} \to 0$.
Therefore, the variance of a quantum field, $\braket{\phi^2(x)}$, is formally infinite.
This highlights the fact that quantum fields are distributions that cannot be multiplied arbitrarily~\cite{Schwartz1954}.
To illustrate this, let us follow~\cite{Banks:2014twn} and consider the correlation function $\braket{\phi(y_1) \ldots \phi(y_r) \phi^n(x)}$, which in terms of Feynman diagrams is given by all graphs featuring $r$ external vertices and one internal vertex with $n$ edges.
Some of these diagrams include loops connecting the internal vertex to itself, leading to divergent loop integrals.
In the free theory, normal ordering effectively removes these divergent contributions, thereby defining the renormalised composite operator $\phi^n(x)$ in the quantum theory.
In an interacting quantum theory, a rigorous regularisation and renormalisation involves the mixing of different operators.
Nevertheless, this procedure is essential for providing a meaningful interpretation of composite operators in QFTs~\cite{Hollands:2023txn}.
To address the divergent nature of these operators, we can formally treat quantum fields as operator-valued distributions that map test functions to operators acting on the Hilbert space of states.
These test functions serve as regulators for the UV behaviour of correlation functions in the short-distance limit, a point we will elaborate on in the following.

\subsection{Quantum fields as operator-valued distributions}

In a scenario where all spacetime points of local field operator insertions coincide, correlation functions are ill defined and their divergent structure needs to be carefully examined.
For instance, this can be done in axiomatic formulation of QFT, where quantum fields are not local operators; rather, they are operator-valued distributions~\cite{Wightman:1956zz, StreaterWightman2001}.
These distributions are linear functionals that map test functions to operators that act on the Hilbert space of states,
\begin{equation}
    \phi_f = \int \md^d x \, \phi(x) f(x) \, .
\end{equation}
Here, $\phi(x)$ is the quantum field understood as an operator-valued distribution, $f$ is a local test function, and $\phi_f$ is the corresponding operator acting on the Hilbert space.
We require the test functions to be smooth and either have compact support in spacetime or are rapidly decreasing as $\absl{x} \to \infty$, along with their derivatives.
Both scenarios provide the field with localised support in spacetime, effectively introducing a UV cutoff for the quantum theory.\footnote{Quantum fields involving test functions with compact support are called \emph{strictly localisable}, while those that are rapidly decreasing are referred to as \emph{tempered distributions}. The spectral density of tempered quantum fields does not grow faster than a fixed-order polynomial, whereas that of strictly localisable operators can grow exponentially with energy~\cite{Jaffe:1967nb}, such as in Galileon theories (see, e.g., \cite{Keltner:2015xda}). For further details and phenomenological consequences, we refer the reader to~\cite{Khoze:2017tjt, Khoze:2017lft, Khoze:2017uga, Khoze:2018bwa, Belyaev:2018mtd, Monin:2018cbi, Khoze:2018qhz}.}
A simple prototype of this concept is realised by smearing the quantum field with a Gaussian test function (see also~\cite{Epstein1973Role}),
\begin{equation}
    f_\sigma (x, t) = \frac{\delta(t)}{\left(2 \pi \sigma^2\right)^{\frac{d-1}{2}}} \exp \left( - \frac{\abs{x}^2}{2 \sigma^2} \right) \, ,
    \label{eq:TestfunctionGaussian}
\end{equation}
where $\sigma$ defines its width.
This choice of test function localises the operator around the origin, approximately over a distance~$\sigma$ in each spatial direction.

A rigorous treatment of quantum fields as operator-valued distribution allows us to assign a meaningful interpretation to the variance of a quantum field, $\braket{\phi^2(x)}$.
The localised support in spacetime, characterised by the width $\sigma$, effectively introduces a UV cutoff to the quantum theory, which regulates the short-distance behaviour of correlation functions.
Naively, the inverse width $\sigma^{-1}$ determines the energy scale beyond which two spacetime points become indistinguishable.
To illustrate this, let us consider a free real scalar field of mass $m$, which is decomposed into momentum modes,
\begin{equation}
    \phi(x) = \int \frac{\md^{d-1} p}{\left( 2\pi \right)^{d-1}} \frac{1}{\sqrt{2 E_p}} \left(a_{\vec{p}} \me^{-ipx} + a_{\vec{p}}^{\dagger} \me^{ipx} \right) \, .
\end{equation}
Here, $a_{\vec{p}}^{\dagger}$ and $a_{\vec{p}}$ are the standard creation and annihilation operators of momentum $\vec{p}$.
By employing the canonical commutation relations for $a_{\vec{p}}^{\dagger}$ and $a_{\vec{p}}$, we can determine the associated operator after smearing it with the Gaussian test function~\eqref{eq:TestfunctionGaussian},
\begin{equation}
    \phi_{f_{\sigma}} = \int \frac{\md^{d-1} p}{\left( 2\pi \right)^{d-1}} \frac{1}{\sqrt{2 E_p}} \me^{-\frac{\sigma^2 \vec{p}^2}{2}} \left( a_{\vec{p}} + a_{\vec{p}}^{\dagger} \right) \, .
\end{equation}
This enables us to define the short-distance limit of the correlation functions by the asymptotic relation
\begin{equation}
	\braket{\phi(x) \phi(y)} \sim \braket{\phi_{f_{\sigma}}^2} \, ,
\end{equation}
as $\absl{x-y} \to 0$.
Similarly, we can express correlation functions at coinciding spacetime points involving $2n$ fields as
\begin{equation}
	\braket{\phi_{f_{\sigma}}^{2n}} = (2n-1)!! I^n \, ,
	\label{eq:MomentsFreeField}
\end{equation}
where we have defined the momentum integral
\begin{equation}
	I = \int \frac{\md^{d-1} p}{\left(2\pi\right)^{d-1}} \frac{1}{2 E_p} \me^{-\sigma^2 \vec{p}^2} \, .
\end{equation}
Evaluating $I$ in spherical coordinates yields Tricomi's confluent hypergeometric function,
\begin{equation}
	I \propto U \left( \frac{1}{2}, 2 - \frac{d}{2}, m^2 \sigma^2 \right) \, .
\end{equation}
Notably, the localised support of the quantum field parametrised by the width $\sigma$, only shows up in combination with its mass, $m \sigma$.
Therefore, since the case $\sigma = 0$ does not belong to the space of smooth test functions, any finite $\sigma$ serves to regulate the theory in the UV.

Intuitively, the short-distance limit of correlation functions in \cref{eq:MomentsFreeField} can be interpreted as the moments of the probability density defined by the classical action.
These moments include disconnected contributions, which in the free theory factorise into products of propagators.
To determine the connected correlation functions at coinciding points, which correspond to the cumulants of the probability density, we can apply \cref{eq:ConnectedGnFaaDiBruno}.
While the moments exhibit factorial growth, we find that the cumulants for a free theory vanish, except for the propagator at $n=1$.
This matches our expectation from Feynman diagrams, as there are no higher-point graphs for a free field with fully-connected topologies.
In fact, the moments and cumulants are represented by Feynman diagrams with a vertex with $n$ edges, which corresponds to the local insertion of the composite operator.
When we include self-interactions, these graphs are similar to petal diagrams where the edges of the vertex are fully-connected for the cumulants, while for the moments the edges can be connected to each other in separate subgraphs.
Let us now investigate their nonperturbative behaviour in quantum theories with nontrivial self-interactions.
To build intuition, we first consider toy models that resemble quantum theories in zero dimensions.

\section{Analogies with Ordinary Integrals}
\label{sec:OrdinaryIntegrals}

Ordinary integrals serve as a valuable test bed for exploring path integrals in the context of interacting QFTs.
They often permit exact solutions while effectively capturing the essential features of QFTs in higher dimensions.
In this toy model, we aim to illustrate the behaviour of correlation functions involving a large number of fields.
We begin by considering the moments $\gamma_n$.
These correspond to the nonconnected correlation functions in zero dimensions and can be expressed as an ordinary integral,
\begin{equation}
	\gamma_n = \int_{-\infty}^{\infty} \md \phi \, \phi^n \exp \left(-\frac{1}{2} \phi^2 - \frac{1}{2p} g^{2(p-1)} \phi^{2p} \right) \, .
	\label{eq:Gamman0D}
\end{equation}
Here, both $n$ and $p$ are positive integers, and $g^2$ denotes the coupling.
The power of the latter is chosen such that, in any perturbative expansion, it aligns with the loop-counting parameter through the field redefinition, $\varphi = g \phi$.

\begin{figure}[t]
	\centering
	\includegraphics{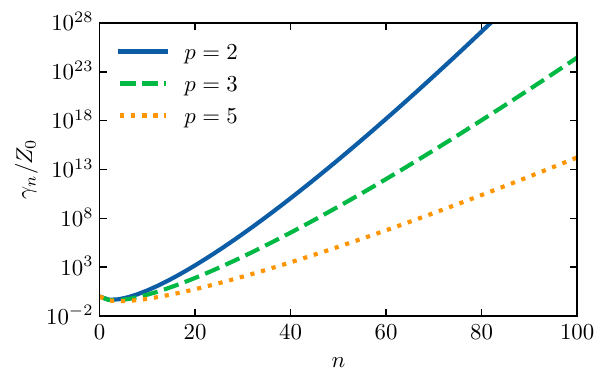}
	\caption{Moments $\gamma_n$ normalised to $Z_0$, as a function of the number of field insertions $n$ in zero dimensions. The different colours represent different self-interaction terms of the theory, $\phi^{2p}$. For simplicity, the coupling is set to one, $g^2=1$.}
	\label{fig:0DGammas}
\end{figure}

In zero dimensions, we can compute the moments $\gamma_n$ in \cref{eq:Gamman0D} exactly for any integer $n$.
The result is illustrated in \cref{fig:0DGammas}, where we show $\gamma_n$ as a function of $n$.
For simplicity, we have set the coupling to $g^2 = 1$.
We find that the moments exhibit factorial growth with increasing $n$.
However, this growth severely depends on the power $p$ of the self-interaction term $\phi^{2p}$.
Clearly, for larger values of $p$ the factorial growth is significantly suppressed at nonperturbatively large $n$.
This behaviour can also be confirmed analytically by neglecting the quadratic terms of the potential, leading to the asymptotic form
\begin{equation}
    \gamma_n \sim \Gamma \left( \frac{n}{2p} \right) \, .
    \label{eq:gammanLargen0D}
\end{equation}
This demonstrates that the factorial growth is drastically reduced by the higher-order self-interactions of the field.

\begin{figure}[t]
	\centering
	\includegraphics{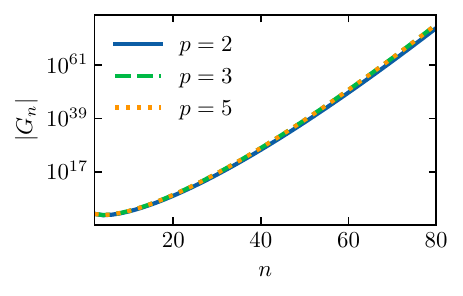}
	\includegraphics{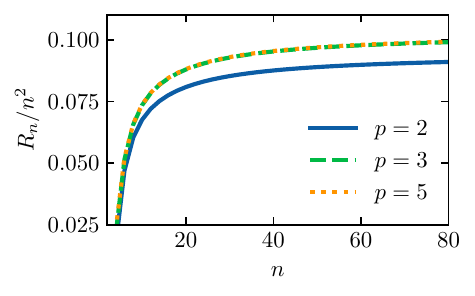}
	\caption{Cumulants $G_n$ (left) and their rescaled ratios (right), as a function of the number of fields $n$ in zero dimensions. The different colours represent different self-interaction terms of the theory, $\phi^{2p}$. For simplicity, the coupling is set to one, $g^2=1$.}
	\label{fig:0DGns}
\end{figure}

We now turn our attention to the cumulants $G_n$, which correspond to the connected correlation functions in zero dimensions.
These can be obtained by substituting the exact expression for the moments $\gamma_n$ into \cref{eq:ConnectedGnFaaDiBruno}.
An example of the cumulants is presented in the left panel of \cref{fig:0DGns}.
We again observe a rapid factorial growth.
However, unlike $\gamma_n$, this growth appears to be largely independent of the specific value of $p$, at least at the factorial level.
Furthermore, the cumulants grow faster than the moments, even for the smallest possible values of $p$.
This observation is both surprising and counterintuitive, as the connected correlation functions are naively derived from the nonconnected ones by subtracting all factorisable diagrams.
Similar phenomena have been noted in other contexts using Dyson-Schwinger equations~\cite{Bender:2022eze, Bender:2023ttu}.

To establish the universal growth of the cumulants at large $n$ more precisely, it is useful to consider their ratio,
\begin{equation}
	R_n = \frac{G_{n+2}}{G_n} \, .
	\label{eq:RatioGn0D}
\end{equation}
If we assume that the cumulants grow factorially, $G_n \propto n!$, then the ratio will be a quadratic function of $n$, $R_n \propto n^2$.
This ratio is illustrated in the right panel of \cref{fig:0DGns} for various values of $p$.
Indeed, we find that schematically
\begin{equation}
	R_n \sim c_p n^2 \, ,
\end{equation}
where the constant $c_p$ exhibits only moderate dependence on $p$.
Substituting this result into \cref{eq:RatioGn0D} allows us to obtain the large-$n$ behaviour of the cumulants,
\begin{equation}
	G_n \sim 2^{n-1} c_p^{\frac{n}{2}-1} \Gamma \left(\frac{n}{2}\right)^2 \, .
\end{equation}
This illustrates that any dependence on the self-interaction term contributing to the cumulants $G_n$ is at most a power law, which is markedly different from the reduced factorial growth observed for the moments $\gamma_n$.
Consequently, we conclude that the factorial growth of the cumulants is universal with respect to $n$, i.e.~mostly independent of the self-interaction term of the theory.
This suggests that, in zero dimensions, the moments $\gamma_n$ are primarily governed by the dynamics of the theory, while the corresponding cumulants $G_n$ are predominantly governed by combinatorial factors.

We note that our results align with established theorems from complex analysis.
In fact, in zero dimensions, we can formally proof that the cumulants grow factorially, without relying on saddle point approximations or the successive application of the chain rule as was done to derive \cref{eq:ConnectedGnFaaDiBruno}.
Closely following~\cite{Goldberg:1990ys}, let us first assume that the generating function for the cumulants (which in zero dimensions is a function rather than a functional) has a formal series expansion with coefficients $a_n$,
\begin{equation}
	\ln Z(J) = \sum_{n=0}^{\infty} a_n J^n \, ,
	\label{eq:0DlnZSeries}
\end{equation}
where $J$ is a complex source.
Suppose that this series representation has a finite radius of convergence, $\absl{J_c}$.
According to Abel's theorem, the asymptotic growth of the coefficients is governed by inverse powers of this radius,
\begin{equation}
	a_n \sim \left(\frac{1}{\abs{J_c}}\right)^n \, ,
\end{equation}
as $n \to \infty$.
Consequently, the cumulants necessarily exhibit factorial growth in this regime,
\begin{equation}
	G_n \sim \left(\frac{1}{\abs{J_c}}\right)^n n! \, .
\end{equation}
Thus, if the series expansion of the generating function in \cref{eq:0DlnZSeries} has a finite radius of convergence, the factorial growth of $G_n$ is a direct consequence.
In fact, $\ln Z(J)$ possesses a finite radius of convergence if the partition function $Z(J)$, as an analytic function, has a root somewhere in the complex plane.
Recall that $Z(J)$ is the generating function of the moments,
\begin{equation}
	Z(J) = \sum_{n=0}^{\infty} \frac{\gamma_n}{n!} J^n \, ,
\end{equation}
where, according to \cref{eq:gammanLargen0D}, the coefficients grow as $\gamma_n \sim \Gamma [ n / (2p) ]$ for large $n$.
This leads to the asymptotic behaviour of the partition function in the complex plane,
\begin{equation}
	Z(J) \sim \exp \left( \abs{J}^{\frac{2p}{2p-1}} \right) \, ,
\end{equation}
as $\absl{J} \to \infty$.
Thus, through elementary complex analysis, we find that the analytic function $Z(J)$ is of order $2p / (2p-1)$.
This order is rational for any value of $p \geq 2$, i.e.~for any nontrivial self-interaction.
Consequently, by Picard's theorem, $Z(J)$ must have infinitely many roots in the complex plane (see, e.g., \cite{Markushevich1966Entire}).
This completes our proof that the cumulants, corresponding to the connected correlation functions, grow factorially with $n$.
With this understanding derived from ordinary integrals, let us now proceed to examine higher-dimensional quantum theories.

\section{Moments and Cumulants from Quantum Path Integrals}
\label{sec:QMandQFT}

While ordinary integrals provide a zero-dimensional test bed that captures the main features of a quantum theory, functional integrals used in QM or QFT settings are more involved to evaluate.
Similarly, in zero dimensions the moments of a probability distribution are well defined, while in a QFT they are not and need to be regulated.
According to our discussion in \cref{sec:CorrelationFunctions}, this can be done by treating quantum fields as operator-valued distributions which are smeared by smooth test functions.
In this case, we can formally write the correlation function at coinciding spacetime points as an object localised around the origin,
\begin{equation}
    \gamma_n = \braket{\phi_{f_{\sigma}}^n} \, .
\end{equation}
Schematically, this means that the correlation function, as a distribution, is convoluted with the Gaussian test function~\eqref{eq:TestfunctionGaussian} of width $\sigma$,
\begin{equation}
    \gamma_n \simeq \int \prod_{i=1}^n \md^d x_i \, f_{\sigma} (x_i) \braket{\phi (x_1) \ldots \phi(x_n)} \, .
\end{equation}
In the regime of small width (compared to their relevant mass scale), each field is sharply localised around the origin.
Therefore, in the limit $\sigma \to 0$, the moments asymptotically approach the path integral expression
\begin{equation}
    \gamma_n \sim \int \mathcal{D} \phi \, \phi^n (0) \me^{-S_E [\phi]} \, .
    \label{eq:gamman}
\end{equation}
Here, we have defined the Euclidean action
\begin{equation}
	S_E [\phi] = \int \md^d x \, \left( \frac{1}{2} \left(\partial \phi \right)^2 + \frac{m^2}{2} \phi^2 + \frac{1}{2p} g^{2(p-1)} \phi^{2p} \right) \, .
\end{equation}
This expression is the higher-dimensional generalisation of the ordinary integral in \cref{eq:Gamman0D}.
However, an exact solution for an interacting quantum theory is lacking, such that we have compute $\gamma_n$ using approximation methods.

\subsection{An effective saddle point approximation}

To determine the nonconnected correlation functions, we need to find an appropriate saddle point approximation of \cref{eq:gamman}.
It is convenient to perform a field redefinition $\varphi = g \phi$, such that the moments read
\begin{equation}
	\gamma_n = g^{-n-1} \int \mathcal{D} \varphi \, \varphi^n \exp \left(-\frac{S_E [\varphi]}{g^2}\right) \, .
	\label{eq:GammanQFT}
\end{equation}
In these coordinates, the Euclidean action does not explicitly depend on the coupling of the theory,\footnote{This is a consequence of the fact that the magnitude of a coupling does not have a physical meaning at the classical level (see, e.g., \cite{Coleman:1985rnk}).}
\begin{equation}
	S_E [\varphi] = \int \md^d x \, \left( \frac{1}{2} \left(\partial \varphi \right)^2 + \frac{1}{2} \varphi^2 + \frac{1}{2p} \varphi^{2p} \right) \, .
\end{equation}
In addition, we have normalised all dimensionful quantities to the scalar field mass, setting $m^2=1$.

In practice, modern QFT calculations evaluate the path integral perturbatively, i.e.~as a series expansion around the vacuum, although they often involve only a limited number of local field insertions.
However, this approach may fail as $n$ increases, since the saddle point approximation is typically performed around the vacuum $\phi = 0$.
Similar to \cref{eq:Gamman0D}, as $n$ grows the support of the integrand shifts away from the perturbative vacuum at the origin and extends towards infinity along the real axis, significantly diminishing the accuracy of the naive approximation.
This limitation of the perturbative approach can also be understood through Feynman diagrams.
Typically, the number of Feynman diagrams contributing to the $n$-th order of the perturbative expansion grows factorially, $c_n \sim n!$, highlighting the asymptotic nature of perturbation theory.
Consequently, for large $n$, the combination of $n$ and coupling, $g^2 n$, distinguishes the perturbative regime, where $g^2 n \ll 1$, from the nonperturbative regime of the theory, where $g^2 n \gtrsim 1$.
This implies that a naive summation of Feynman diagrams in the latter regime does not yield an accurate and meaningful result for large $g^2 n$.
To remedy this issue, we instead follow a semiclassical approach and express the moments as (see also~\cite{Badel:2019oxl, Badel:2019khk})
\begin{equation}
	\gamma_n = g^{-n-1} \int \mathcal{D} \varphi \, \exp \left(-\frac{S_{\mathrm{eff}} [\varphi]}{g^2}\right) \, ,
\end{equation}
with an effective action that accounts for the local operator insertion,\footnote{Due to parity symmetry, we can disregard any imaginary part arising from the logarithmic term due to negative field values, without loss of generality.}
\begin{equation}
	S_{\mathrm{eff}} [\varphi] = S_E [\varphi] - \epsilon \log \varphi (0) \, .
\end{equation}
Here, for convenience, we have defined the combination
\begin{equation}
	\epsilon = g^2 n \, ,
\end{equation}
which allows us to systematically construct a double scaling limit where $g^2 \to 0$ and $n \to \infty$ while keeping $\epsilon$ fixed.
In this case, $\epsilon$ captures the dynamics relevant to the weakly-coupled quantum theory.
As previously mentioned, $\epsilon \ll 1$ characterises the perturbative regime, while $\epsilon \gtrsim 1$ characterises the nonperturbative domain of the theory.
In other words, perturbation theory yields reliable and accurate results for small $\epsilon$, but breaks down for large $\epsilon$.

Minimising the effective action $S_{\mathrm{eff}}$ accounts for the shifted support of the integrand away from the perturbative vacuum towards infinity as $n$ increases.
Thus, we can construct a suitable saddle point approximation around this nontrivial field configuration, satisfying $\delta S_{\mathrm{eff}} / \delta \varphi = 0$.
The corresponding equations of motion for $\varphi$ are given by
\begin{equation}
	\partial^2 \varphi - \varphi - \varphi^{2p-1} = -\epsilon \frac{\delta^{(d)}(x)}{\varphi(0)} \, .
	\label{eq:eom}
\end{equation}
Here, the local field insertion acts as an inhomogeneous source term for the field's dynamics, which therefore explicitly depend on the effective coupling $\epsilon$.
Since there are no known analytic solutions to \cref{eq:eom}, we make a perturbative ansatz to find the classical field configuration corresponding to the effective saddle point,
\begin{equation}
	\varphi(x) = \sqrt{\epsilon} \left( \varphi_0(x) + \epsilon^{p-1} \varphi_1(x) + \epsilon^{2(p-1)} \varphi_2(x) + \ldots \right) \, .
	\label{eq:FieldAnsatz}
\end{equation}
Using this ansatz, we can systematically solve the dynamics order by order in powers of $\epsilon$ by determining the coefficient functions $\varphi_i(x)$.
For instance, at leading order, the equations of motions take the form
\begin{equation}
	\partial^2 \varphi_0 - \varphi_0 = - \frac{\delta^{(d)}(x)}{\varphi_0(0)} \, .
	\label{eq:EomPhi0}
\end{equation}
Thus, $\varphi_0$ is proportional to the Green's function of the Helmholtz operator in $d$ dimensions, $\varphi_0 \propto r^{-d/2+1} K_{d/2-1}(r)$.
Similarly, the effective action can be decomposed into a power series in $\epsilon$ through \cref{eq:FieldAnsatz}.
At leading order, it involves only the field $\varphi_0$ and is given by
\begin{equation}
	S_{\mathrm{eff},0} = \epsilon \int \md^d x \, \left( \frac{1}{2} \left(\partial \varphi_0\right)^2 + \frac{1}{2} \varphi_0^2 \right) - \epsilon \log \left[ \sqrt{\epsilon} \varphi_0 (0) \right]
	= \frac{\epsilon}{2} - \frac{\epsilon}{2} \log \left( \frac{\epsilon}{2} \right) \, .
	\label{eq:Seff0}
\end{equation}
Consequently, the moments at large $n$ schematically read
\begin{equation}
	\gamma_n \sim \exp \left[ \frac{1}{g^2} \left( \frac{\epsilon}{2} \log \left(\frac{\epsilon}{2}\right) - \frac{\epsilon}{2} \right) \right] \, .
	\label{eq:GammanTreeLevel}
\end{equation}
We remark that this result should be understood within the context of a double scaling limit where $g^2 \to 0$ and $n \to \infty$, while keeping $\epsilon$ fixed.
In fact, \cref{eq:GammanTreeLevel} is a large-$n$ representation of the familiar factorial growth of tree-level perturbation theory, through Stirling's approximation.
For example, the exponential form of $\gamma_n$ agrees with the overlap of an off-shell quantum state (created by an off-shell field acting on the vacuum) and an arbitrary $n$-particle state, $\braket{n | \varphi | 0}$.
More precisely, it matches the tree-level computation of the latter in the double scaling limit~\cite{Cornwall:1990hh, Goldberg:1990qk, Brown:1992ay, Voloshin:1992mz, Argyres:1992np}.
Physically, this is expected, as in both scenarios, any tree-level approximation is dominated by the number of Feynman diagrams, which grow factorially, thus yielding the above behaviour through Stirling's approximation (modulo appropriate powers of the coupling).

So far, \cref{eq:FieldAnsatz} dictates a perturbative expansion around $\epsilon = 0$, which we expect to yield an accurate approximation in the perturbative regime of the quantum theory, where $\epsilon \ll 1$.
In principle, it is viable to systematically compute higher-order terms of the classical field configuration, $\varphi_i$, along with the corresponding corrections to the effective action.
This would extend the range of validity of the saddle point approximation.
However, going beyond perturbation theory into the nonperturbative domain, where $\epsilon \gtrsim 1$, requires the resummation of a substantial number of terms in this expansion.
While exploring this domain is feasible in CFTs (see~\cite{Badel:2019oxl, Badel:2019khk, Antipin:2024ekk}), scenarios involving a mass term present additional complications.
For example, although formal solutions for $\varphi_1$ exist, determining the corresponding correction to the effective action in an arbitrary number of dimensions proves to be challenging.
One of the simplest nontrivial examples where an explicit solution can be obtained is QM, which serves as a one-dimensional QFT.
In this context, the first correction to the effective action is given by
\begin{equation}
	S_{\mathrm{eff},1} = \frac{\epsilon^p}{2^{p+1} p^2} \, .
	\label{eq:Seff1QM}
\end{equation}
A detailed derivation of this result is provided in \cref{app:CorrectionSeff}.
Accordingly, we will perform a complete resummation of all higher-order corrections to the saddle point approximation of the effective action in QM.

\subsection{Resummation of the effective action in quantum mechanics}

Addressing the nonperturbative behaviour of correlation functions at short distances requires a resummation of the perturbative expansion of the effective action to all orders.
We expect this approach to yield the exact results, as there are no instantons present on the real axis in the Borel plane that could spoil the resummation.
In QM, this resummation of the effective action is feasible because instead of relying on a path integral formulation, we can compute the moments directly from the ground-state wave function,
\begin{equation}
	\gamma_n = \int_{-\infty}^{\infty} \md x \, x^n \abs{\psi_0(x)}^2 \, .
	\label{eq:GammanQM}
\end{equation}
The wave functions, in turn, can be systematically derived from a perturbative expansion of Schr\"odinger's equation at weak coupling,
\begin{equation}
	\left(-\frac{1}{2} \frac{\md^2}{\md x^2} + \frac{1}{2} x^2 + \frac{1}{2p} g^{2(p-1)} x^{2p} - E \right) \psi (x) = 0 \, .
	\label{eq:schroedinger}
\end{equation}
We make an ansatz for both the wave functions as well as the corresponding energy levels as a formal series expansion in powers of $g^{2(p-1)}$, allowing us to solve \cref{eq:schroedinger} order by order in this expansion.
More generally, the perturbative construction of the spectrum is performed through a set of recursion relations introduced by Bender and Wu~\cite{Bender:1969si, Bender:1973rz}.
To implement the recursion relations in practice, we use the efficient code \texttt{BenderWu}~\cite{Sulejmanpasic:2016fwr}.

For example, for the standard quartic anharmonic oscillator with $p=2$, we obtain the ground-state wave function
\begin{equation}
	\psi_0^{p=2}(x) = \me^{-\frac{x^2}{2}} \left[1 - \frac{g^2}{16} \left(x^2 + 3\right) + \frac{g^4}{512} \left(x^8 + \frac{26}{3} x^6 + 31 x^4 + 84 x^2 \right) + \ldots \right] \, .
\end{equation}
Here, we have normalised the wave function such that $\int_{-\infty}^{\infty} \md x \, \abs{\psi_0}^2 = 1$.

Using Bender's and Wu's method, we can compute the spectrum of \cref{eq:schroedinger} to any desired order in the perturbative expansion.
This, in turn, determines the moment in \cref{eq:GammanQM} up to the same perturbative order.
Notably, we find that each series coefficient of $\gamma_n$ is a polynomial in $n$.
For instance, for $p=2$, the first few terms take the form
\begin{equation}
	\gamma_n^{p=2} = \frac{\Gamma\left(\frac{n+1}{2} \right)}{\sqrt{\pi}} \left[ 1
	- \frac{g^2}{32} \left( n^2 + 10n \right)
	+ \frac{g^4}{2048} \left( n^4 + \frac{92}{3} n^3 + 280 n^2 + \frac{2968}{3} n \right) + \ldots \right] \, ,
	\label{eq:GammanP2}
\end{equation}
while for $p=3$ they read
\begin{equation}
	\begin{split}
		\gamma_n^{p=3} =& \frac{\Gamma \left(\frac{n+1}{2} \right)}{\sqrt{\pi}}
		\left[ 1 - \frac{g^4}{144} \left( n^3 + \frac{33}{2} n^2 + 98 n \right) \right. \\
		&+ \left. \frac{g^8}{41472} \left( n^6 + \frac{273}{5} n^5 + \frac{4807}{4} n^4 + 13104 n^3 + 75079 n^2 + \frac{1136382}{5} n \right) + \ldots \right] \, .
	\end{split}
	\label{eq:GammanP3}
\end{equation}
In practice, we compute hundreds of orders of this formal series expansion and identify patterns in their structure.
Regardless of the interaction's power $p$, we find a universal behaviour of the leading-order (tree level) contribution,
\begin{equation}
	\gamma_n^{\mathrm{tree}} = \frac{\Gamma \left( \frac{n+1}{2} \right)}{\sqrt{\pi}} \, .
\end{equation}
We have confirmed this result for self-interactions up to $p=10$.
It exemplifies the familiar factorial growth of observables associated with high multiplicities or large quantum numbers, highlighting the asymptotic nature of perturbation theory.
Furthermore, we find that this leading contribution factorises from the full expression, as indicated in \cref{eq:GammanP2,eq:GammanP3}.
This suggests that the remaining loop contributions can be effectively resummed in the large-$n$ limit, similar to the case of the QM analogue of high-multiplicity form factors, $\braket{n | x | 0}$~\cite{Jaeckel:2018ipq, Jaeckel:2018tdj, LoChiatto:2023bam}.
The leading quantum corrections to the tree-level result indicate an exponential form,
\begin{alignat}{2}
	\gamma_n^{p=2} &\sim 1 - \frac{g^2 n^2}{32} + \frac{g^4 n^4}{2048} - \frac{g^6 n^6}{196608} + \ldots &&= \exp \left(- \frac{g^2 n^2}{32} \right) \, , \\
	\gamma_n^{p=3} &\sim 1 - \frac{g^4 n^3}{144} + \frac{g^8 n^6}{41472} - \frac{g^{12} n^9}{17915904} + \ldots &&= \exp \left(- \frac{g^4 n^3}{144} \right) \, .
\end{alignat}
The exponentiation is crucial for exploring the behaviour of the correlation functions in the double scaling limit $g^2 \to 0$ and $n \to \infty$, while keeping their combination $\epsilon = g^2 n$ fixed.
Specifically, if we adopt the general ansatz
\begin{equation}
	\gamma_n = \frac{\Gamma \left(\frac{n+1}{2} \right)}{\sqrt{\pi}} \exp \left( \frac{F_c}{g^2} \right) \, ,
	\label{eq:GammanAnsatz}
\end{equation}
the large-$n$ limit of $\gamma_n$ is entirely encoded in both the tree-level contribution and, more importantly, in the exponent function $F_c$.
If $F_c$ is negative and sufficiently large to counteract the tree-level growth, one may conclude that $\gamma_n$ is exponentially suppressed at high energies.

To shed some light on this behaviour, we note that $F_c$ can be systematically constructed in an asymptotic $1/n$-expansion (see also~\cite{Jaeckel:2018ipq, Jaeckel:2018tdj}),
\begin{equation}
	F_c(\epsilon, n) = F_0(\epsilon) + \frac{1}{n} F_1(\epsilon) + \frac{1}{n^2} F_2(\epsilon) + \ldots \, .
	\label{eq:FSchematic}
\end{equation}
This implies that in the limit where $\epsilon$ is kept fixed, the large-$n$ behaviour of $F_c (\epsilon)$ is dominated by the contribution $F_0 (\epsilon)$.
This holds true both in the perturbative, where $\epsilon \ll 1$, as well as in the nonperturbative regime, where $\epsilon \gtrsim 1$.

Let us now determine the exponent $F_c$ by expressing its coefficient functions in \cref{eq:FSchematic} as
\begin{equation}
	F_i(\epsilon) = \epsilon^p \sum_{k=0}^{\infty} c_{i, k} \epsilon^{k(p-1)} \, .
	\label{eq:FCoeffsAnsatz}
\end{equation}
We can then fix all coefficients $c_{i,k}$ by substituting this ansatz into \cref{eq:GammanAnsatz}, performing the series expansion for the exponential, and matching the result to the analytic expressions for $\gamma_n$, such as those given in \cref{eq:GammanP2,eq:GammanP3}.
For the first few terms, we find
\begin{equation}
	\begin{split}
		F_c^{p=2} =& -\frac{\epsilon^2}{32} + \frac{\epsilon^3}{192} - \frac{35}{24576} \epsilon^4 + \ldots
		+ \frac{1}{n} \left( - \frac{5}{16} \epsilon^2 + \frac{45}{512} \epsilon^3 + \ldots \right) \\
		&+ \frac{1}{n^2} \left( \frac{371}{768} \epsilon^2 - \frac{3913}{12288} \epsilon^3 + \ldots \right)
		+ \mathcal{O}\left(\frac{1}{n^3}\right) \, ,
	\end{split}
\end{equation}
and, for $p=3$,
\begin{equation}
	\begin{split}
		F_c^{p=3} =& -\frac{\epsilon^3}{144} + \frac{\epsilon^5}{1920} - \frac{5}{72576} \epsilon^7 + \ldots
		+ \frac{1}{n} \left( - \frac{11}{96} \epsilon^3 + \frac{163}{9216} \epsilon^5 + \ldots \right) \\
		&+ \frac{1}{n^2} \left( - \frac{49}{72} \epsilon^3 + \frac{1645}{6912} \epsilon^5 + \ldots \right)
		+ \mathcal{O}\left(\frac{1}{n^3}\right) \, .
	\end{split}
\end{equation}
To determine the large-$n$ behaviour of the moments in the double scaling limit described above, we now write
\begin{equation}
	\gamma_n = \frac{1}{g^{2n}} \exp \left( \frac{F}{g^2} \right) \, ,
	\label{eq:Gamman1DGeneral}
\end{equation}
such that we include the tree-level factor in the exponent using Stirling's formula,
\begin{equation}
	F(\epsilon, n) = \frac{\epsilon}{2} \left( \log \frac{\epsilon}{2} - 1 \right) + F_c(\epsilon, n) \, .
\end{equation}
It is remarkable that this exponentiation is exact to all orders in perturbation theory, which is not guaranteed (for a detailed discussion of exponential series representations, see~\cite{Jaeckel:2018ipq}).

\begin{figure}[t]
	\centering
	\includegraphics{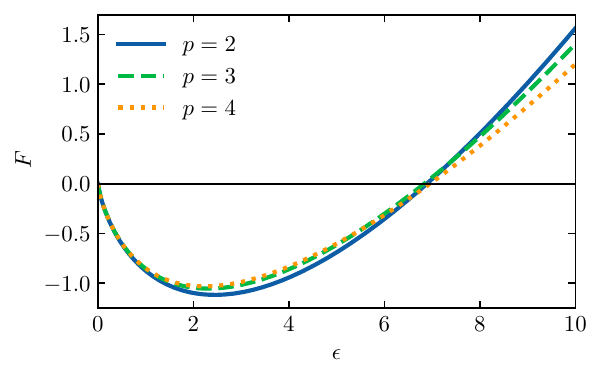}
	\caption{Resummed exponent $F$ of the moments $\gamma_n$, as a function of $\epsilon$ in the double scaling limit $g^2 \to 0$ and $n \to \infty$, where $\epsilon = g^2 n$ is kept fixed. The different colours represent different self-interaction terms of the theory, $\phi^{2p}$.}
	\label{fig:1dHolygrail}
\end{figure}

By construction, this result is an expansion about the perturbative regime at $\epsilon = 0$.
Therefore, exploring the nonperturbative domain of the theory, where $\epsilon \gtrsim 1$, requires a suitable resummation of $F$ beyond the regime where $\epsilon$ is small.
In principle, a resummation procedure based on a finite number of terms is provided by the Borel-Pad\'e approximation (for an introduction to this method, see, e.g., \cite{Bender2013Advanced}).
However, in our case, a close inspection of a large number of terms reveals a closed-form expression for the corresponding coefficients $c_{0,k}$ in \cref{eq:FCoeffsAnsatz} which is valid for arbitrary $p$,
\begin{equation}
	c_{0,k} = \frac{\left(-1\right)^{k+1}}{4} \frac{\left(2^{p-1} p \right)^{-k-1}}{\left(p-1\right) k + p} \frac{\Gamma \left[ \frac{(p+1)(k+1)}{2} \right]}{\Gamma \left[ \frac{(p-1)k + p+1}{2}\right] \Gamma \left( k+2 \right)} \, .
	\label{eq:c0kClosed}
\end{equation}
For instance, we find that the leading-order coefficient of the exponent beyond tree level is given by
\begin{equation}
	c_{0,0} = -\frac{1}{2^{p+1} p^2} \, .
\end{equation}
This expression precisely matches the leading-order correction to the effective action $S_{\mathrm{eff}}$ in \cref{eq:Seff1QM}, which we have obtained from an effective saddle point approximation.
This not only validates our approach to some extent but also facilitates an immediate identification of the exponent function $F_0$ with the effective saddle point.
In this context, it would be interesting to investigate whether the $1/n$-corrections of $F$ correspond to the quantum fluctuations around the effective saddle point.
At the same time, our result demonstrates that each coefficient of $S_{\mathrm{eff}}$ (or equivalently $F_0$) effectively resums an infinite number of loops at large $n$.
This also means that the closed form in \cref{eq:c0kClosed} enables us to resum the formal series expansion and obtain the analytic continuation of $F_0(\epsilon)$, which is valid everywhere in the complex $\epsilon$-plane.
We omit its explicit form here, as it is lengthy and convoluted.
Instead, we illustrate its shape in \cref{fig:1dHolygrail}.

Clearly, $F$ exhibits a turning point before becoming positive beyond a certain critical value $\epsilon_0$, where $F(\epsilon_0) = 0$.
The turning point is reminiscent of the point of optimal truncation, $\epsilon_{\mathrm{opt}}$.
It marks the order beyond which the individual terms of the asymptotic series expansion of $F$ start to grow rapidly, and a careful resummation becomes necessary.
We find that its location only mildly depends on the self-interaction power, and it appears to asymptotically approach $\epsilon_{\mathrm{opt}} \simeq 2$ as $p$ grows.
We have verified this behaviour for self-interactions up to $p=8$.
Similarly, the root of $F$ marks the point $\epsilon_0$ where exponential growth for the moments $\gamma_n$ sets in.
That is, if one chooses a weak coupling $g^2$, $\epsilon_0$ indicates the value of $n$ for which the moments start to grow exponentially, $n \gtrsim \epsilon_0 / g^2$.
While, to the best of our knowledge, an analytic expression for $\epsilon_0$ does not exist for any value of $p$, we observe that it is monotonically shifted towards larger values as $p$ grows.
Again, we have explicitly verified this for self-interactions up to $p=8$.
We expect that it asymptotically behaves as $\epsilon_0 \to \infty$ as $p \to \infty$, as this scenario naively corresponds to the infinite square well potential.
In this case, one can show explicitly that the moments are exponentially suppressed for large $n$, such that $F$ is necessarily negative everywhere.
We conclude that the growth rate depends on the self-interaction power, with larger values of $p$ leading to a suppression of this growth.
This behaviour is consistent with our findings for ordinary integrals (see~\cref{fig:0DGammas}).
Focussing on the nonperturbative domain, where $\epsilon \gtrsim 1$, this observation is further supported by an asymptotic expansion as $\epsilon \to \infty$,
\begin{equation}
	F(\epsilon) \sim \frac{\epsilon}{p+1} \left[ \log \left( \sqrt{\frac{p}{4}} \epsilon \right) - 1 \right] \, ,
	\label{eq:1DHolygrailNonperturbative}
\end{equation}
which clearly shows a suppression factor for large $p$.
In other words, at high energies, low-point interactions seem to dominate $n$-particle rates.

In summary, we find that in the QM analogue of self-interacting QFTs, the moments $\gamma_n$ exponentiate and grow rapidly with a large number of field insertions at high energies, $n \to \infty$.
However, this rapid growth is suppressed by the power $p$ of the field's self-interaction.
In the following, we will argue that this behaviour is in strong contrast to that observed for the cumulants, which correspond to fully connected correlation functions.

\subsection{Asymptotic behaviour of cumulants}

In general, having derived the resummed expression for the moments $\gamma_n$, the corresponding cumulants $G_n$, which represent connected correlation functions of fields inserted at a single spacetime point, can be read off from \cref{eq:ConnectedGnFaaDiBruno}.
Note that we have also normalised the wave functions to $\int_{-\infty}^{\infty} \md x \, \absl{\psi_0}^2 = 1$, which corresponds to the normalisation $Z_0 = 1$.

The computation of the cumulants has to be treated with some care.
Formally, we have determined the analytic continuation of $\gamma_n$ in \cref{eq:Gamman1DGeneral} in the double scaling limit $g^2 \to 0$ and $n \to \infty$, while keeping $\epsilon = g^2 n$ fixed.
However, this limit is not inherently applicable to the expression for $G_n$ given in \cref{eq:ConnectedGnFaaDiBruno}.
Therefore, in practice, we replace any power of the coupling constant by the corresponding expression involving $\epsilon$ as the effective coupling, $g^2 = \epsilon / n$.
We then select an arbitrary but fixed value of $\epsilon$, for instance, in the nonperturbative regime $\epsilon \gtrsim 1$, and subsequently take the large-$n$ limit.
In other words, to obtain a meaningful expression as $n \to \infty$, we express the cumulants as functions of both $n$ and $\epsilon$.
Furthermore, we can organise any subleading corrections to $G_n$ as a series expansion in inverse powers of $n$.

\begin{figure}[t]
	\centering
	\includegraphics{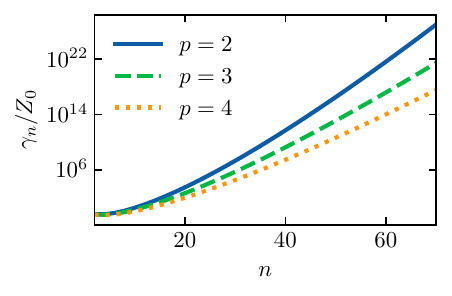}
	\includegraphics{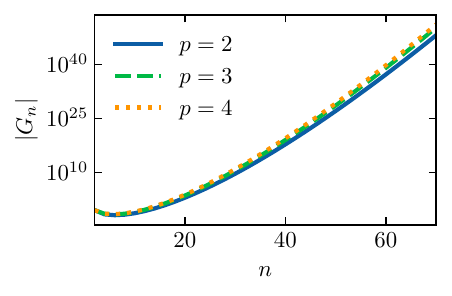}
	\caption{Normalised moments $\gamma_n$ (left) and cumulants $G_n$ (right), as a function of $n$ in one dimension. The different colours represent the self-interaction terms of the theory, $\phi^{2p}$. For simplicity, both $\gamma_n$ and $G_n$ are considered at $g^2 = 1$.}
	\label{fig:1DDisconnectedConnected}
\end{figure}

While we can study the cumulants' behaviour for any value of $\epsilon$ in this manner, the evaluation of \cref{eq:ConnectedGnFaaDiBruno} becomes numerically unstable for large values of $\epsilon$ due to the combinatorial complexities arising from the addition and subtraction of large numbers.
To address this instability, we numerically determine the ground state wave function with high precision and then directly integrate the correlation functions $\gamma_n$, for arbitrary values of $n$ and $g^2$.
These agree with our perturbative results and yield robust results for the cumulants $G_n$ determined from \cref{eq:ConnectedGnFaaDiBruno}.
An example of both $\gamma_n$ and $G_n$ is illustrated in \cref{fig:1DDisconnectedConnected}.
Notably, the moments $\gamma_n$, shown in the left panel of \cref{fig:1DDisconnectedConnected}, exhibit a strong dependence on the self-interaction term of the theory as $n \to \infty$.
This behaviour aligns with our expectations in the nonperturbative regime, where we have established a suppression of the exponent $F$ through the self-interaction power $p$, given in \cref{eq:1DHolygrailNonperturbative}.
In contrast, the cumulants $G_n$ are largely independent of the self-interaction term, meaning they do not significantly depend on $p$.
More precisely, as $n \to \infty$, they grow asymptotically as
\begin{equation}
	G_n \sim 2^{n-1} c_p^{\frac{n}{2}-1} \Gamma \left(\frac{n}{2}\right)^2 \, .
\end{equation}
This behaviour is consistent with the analysis of ordinary integrals presented in \cref{sec:OrdinaryIntegrals}.
We again find that the cumulants grow faster with the number of field insertions than their counterparts that include disconnected contributions.
This observation is somewhat counterintuitive, as all disconnected contributions are subtracted from the correlation functions to obtain the fully connected ones.

Extending these results to higher-dimensional QFTs represents an intriguing avenue for future research.
One potential approach is to systematically compute the series representation of the effective action in powers of $\epsilon$, as outlined earlier.
In the case of QM, we have shown that by comparing the effective action to the exponent function $F$, this method appears to yield reliable results and is indeed equivalent to a large-order perturbative analysis of Schr\"odinger's equation.
However, the expansion of the effective action in higher-dimensional QFTs is inherently more complex.
Unfortunately, given that an all-orders resummation is required, achieving this may not be feasible.
An alternative approach could involve employing fully nonperturbative Dyson-Schwinger equations, as explored in~\cite{Bender:2022eze, Bender:2023ttu}.
We leave this for future work.

That said, it would be interesting to investigate the massless regime of our results, naively resembling the CFT limit of the quantum theory.
In practice, this may be achieved through the framework of ``exact perturbation theory," where the perturbative ansatz for a theory with massive degrees of freedom can be suitably reorganised to obtain nonperturbative results for its massless limit~\cite{Serone:2016qog, Serone:2017nmd}.
While this method, in principle, facilitates the precise computation of vacuum expectation values in the massless limit, their physical interpretation is somewhat more involved.
On a technical level, the ultra-short distance limit of moments in a quantum theory featuring conformal symmetry may not be meaningful, as any test function providing a localised support for the field necessarily introduces a characteristic length scale, beyond which individual spacetime points become indistinguishable.
This length scale breaks the conformal invariance of the quantum theory.
More generally and strictly speaking, a CFT also does not feature the notion of a finite separation between quanta, such that there are no asymptotic in- and out-states to describe scattering processes.
Consequently, connected correlation functions and their associated $S$-matrix elements cannot be properly defined.

On the other hand, the semiclassical expansion around the effective saddle point, as outlined above, is able to shed light on quantum theories featuring global symmetries with an associated Noether charge.
Naively, this Noether charge corresponds to the number $n$ of operator insertions at the same spacetime point.
In this case, the method may facilitate an interpolation between Feynman-diagrammatic calculations, valid for $\epsilon \ll 1$, and the superfluid regime for CFTs at large charge, $\epsilon \gg 1$~\cite{Badel:2019oxl, Badel:2019khk}.
Closely following this approach, it would be interesting to supplement these examples with a massive degree of freedom.
In this case, even determining only the first few terms of the effective action may provide valuable insights into the dynamics of large-charge sectors of the quantum theory.

Finally, our results are intriguing in the context of effective field theories (EFTs), which accommodate operators of arbitrary power beyond renormalisable QFT models.
In this framework, our results suggest that the physical information of the quantum theory carried by the fully connected functions is only marginally dependent on the precise operator structure in the UV.
That is, in the large-$n$ limit, all operators of the EFT contribute equally.
In other words, in the nonperturbative regime of a quantum theory, we would not be able to detect the presence of higher-order interactions through the form factors discussed in this work.
However, higher-dimensional QFTs introduce additional complications that must be carefully addressed.
Notably, all coupling constants beyond a $\phi^4$ term are dimensionful in four dimensions, which introduces an additional energy scale above which higher-dimensional operators become effective.
This must be accounted for in the double scaling limit.
Furthermore, it is important to recognise that the moments and cumulants discussed in this work do not represent physical observables in a strict sense.
In a more realistic setting, physical correlation functions would be measured through scattering processes in particle collisions.
This scenario would involve field insertions at different spacetime points, introducing nontrivial kinematics that can lead to suppression of interaction rates through phase space factors at large $n$, as the quantum theory would otherwise violate unitarity.

\section{Conclusions}
\label{sec:conclusions}

In this work, we have investigated the nonperturbative behaviour of $n$-point correlation functions at short distances, with a particular focus on the role of self-interaction terms $\phi^{2p}$.
In the ultra-short distance limit, where all local field insertions coincide, the nonconnected and connected correlation functions correspond to the moments and cumulants of the probability density defined by the classical action, respectively.
In the context of quantum theory, these objects can be given a well-defined meaning through an axiomatic formulation of QFT, where quantum fields are treated as operator-valued distributions.

To explore the dynamics of these moments and cumulants in the nonperturbative regime, we have adopted a double scaling limit of weak coupling, $g^2 \to 0$, and large quantum number, $n \to \infty$, while keeping the product $g^2 n$ fixed.
We first follow a semiclassical approach, performing an expansion around the saddle points of an effective action that explicitly incorporates the field insertions inside the correlation function.
This approximation is mainly accurate in the regime where the effective coupling is small, $g^2 n \ll 1$.
Therefore, addressing the nonperturbative regime, where $g^2 n \gtrsim 1$, requires enhancing the approximation's range of validity, which can be achieved by a resummation of the perturbative saddle point expansion.
While such computations may be feasible in scenarios featuring a lot of symmetry, they become increasingly challenging in more generic cases.
We therefore perform the resummation of the saddle point expansion to all orders in $g^2 n$ in both zero and one dimensions, serving as a test bed for QFT.
In this case, we derive exact solutions for the moments and cumulants, extending our analysis beyond the perturbative regime.
Our key finding is that the power $p$ of the self-interaction term primarily influences the growth of moments, representing correlation functions that include disconnected contributions.
In contrast, cumulants, representing the fully-connected correlation functions that encapsulate the physical information of the quantum theory, exhibit a universal growth with respect to $n$, largely independent of the specific self-interaction term.
Furthermore, we find that the cumulants grow more rapidly than the moments.

Our results challenge naive expectations, suggesting that the nonperturbative dynamics of quantum theories, encoded in fully-connected correlation functions, exhibit a universal scaling behaviour at large $n$.
This indicates that their dynamics are predominantly governed by the combinatorial aspects of Feynman diagrams rather than the specific dynamics of the self-interactions.
These insights may provide a crucial step towards understanding nonperturbative phenomena in QFTs, particularly in scenarios involving large quantum numbers.

Nevertheless, we note that our analysis is most robust in the context of QM, which can be interpreted as a one-dimensional QFT.
While we anticipate that our main results may extend to higher-dimensional QFTs, a rigorous examination of these theories needs to address additional complications, including nontrivial phase space factors and the careful renormalisation of composite operators.
Future research endeavours should consider these aspects to fully elucidate the nonperturbative behaviour of large-$n$ correlation functions in realistic QFT models.
In addition, investigating the interplay between different operators in effective field theories would provide a deeper understanding of the observed universality and its implications for low-energy phenomenology.
These aspects certainly merit further investigation.

\section*{Acknowledgements}

I am grateful to Carlos Tamarit and Felix Yu for many helpful and interesting discussions on related work.
The author is supported by the Cluster of Excellence \emph{Precision Physics, Fundamental Interactions and Structure of Matter} (PRISMA$^+$ EXC 2118/1) funded by the German Research Foundation (DFG) within the German Excellence Strategy (Project No.~390831469).

\appendix

\section{Perturbative Corrections to the Effective Saddle Point}
\label[appendix]{app:CorrectionSeff}

In a $d$-dimensional Euclidean spacetime, we can define an effective action that accounts for the local operator insertions inside the path integral,
\begin{equation}
	S_{\mathrm{eff}} = \int \md^d x \, \left( \frac{1}{2} \left(\partial \varphi\right)^2 + \frac{1}{2}\varphi^2 + \frac{1}{2p} \varphi^{2p} \right) - \epsilon \log \varphi(0) \, .
\end{equation}
The corresponding equations of motion for the field that minimise the action, $\delta S_{\mathrm{eff}} / \delta \varphi = 0$, are given by
\begin{equation}
	\partial^2 \varphi - \varphi - \varphi^{2p-1} = - \frac{\epsilon}{\varphi(0)} \delta^{(d)}(x) \, ,
\end{equation}
such that the local operator insertions now act as an inhomogeneous source term.
As discussed in the main text, we can systematically solve these equations of motion by employing the ansatz for the field
\begin{equation}
	\varphi(x) = \sqrt{\epsilon} \left( \varphi_0(x) + \epsilon^{p-1} \varphi_1(x) + \epsilon^{2(p-1)} \varphi_2(x) + \ldots \right) \, .
\end{equation}
The leading-order solution to the equations of motion, $\varphi_0$, along with the corresponding effective action, $S_{\mathrm{eff},0}$, are given in \cref{eq:EomPhi0,eq:Seff0}.

Let us now outline how to obtain corrections to the effective action beyond leading order.
Following our perturbative approach, we find the equations of motion for the first correction to the saddle point field configuration, $\varphi_1$, to be
\begin{equation}
	\partial^2 \varphi_1 - \varphi_1 = \varphi_0^{2p-1} + \frac{\varphi_1(0)}{\varphi_0^2(0)} \delta^{(d)}(x) \, .
\end{equation}
Here, all contributions on the right-hand side are known, in principle.
Since the self-interactions of the field appear only as inhomogeneous terms in the equations of motion, we can express a formal solution for $\varphi_1$ using the free Green's functions $D$ in position space,
\begin{equation}
	\varphi_1(x) = \int \md^d y \, D(x,y) \varphi_0^{2p-1}(y) + D(x, 0) \frac{\varphi_1(0)}{\varphi_0^2(0)} \, .
\label{eq:Phi1Formal}
\end{equation}
We note that this contribution to the classical field configuration at the saddle point is of order $\mathcal{O} \left( \epsilon^{p-1} \right)$ with respect to the leading-order result $\varphi_0$.
Similarly, the first correction to the effective action is of order $\mathcal{O}\left(\epsilon^p\right)$ and reads
\begin{equation}
	S_{\mathrm{eff},1} = \epsilon^p \int \md^d x \, \left( \partial \phi_0 \partial \phi_1 + \varphi_0 \varphi_1 + \frac{1}{2p} \varphi_0^{2p} \right) + \ldots \, ,
\label{eq:Seff1Formal}
\end{equation}
where the ellipsis denotes the logarithmic contributions from the local operator insertions.
Although we have formal solutions for both $\varphi_0$ and $\varphi_1$, \cref{eq:Seff1Formal} is difficult to evaluate in general.

In the case of QM, we have $d=1$, simplifying all multidimensional integrals to one-dimensional integrals.
Here, the leading-order contribution to the saddle point configuration is given by
\begin{equation}
	\varphi_0(t) = \frac{1}{\sqrt{2}} \me^{-\abs{t}} \, .
\end{equation}
Furthermore, using the free Green's function
\begin{equation}
	D\left(t, t^{\prime}\right) = \frac{1}{2} \me^{- \abs{t - t^{\prime}}} \, ,
\end{equation}
we can explicitly evaluate the formal expression for $\varphi_1$ in \cref{eq:Phi1Formal}.
This, in turn, allows us to determine $S_{\mathrm{eff},1}$, by evaluating the spacetime integrals in \cref{eq:Seff1Formal} and considering a suitable series expansion in $\epsilon$ of the logarithmic terms, such that we finally arrive at
\begin{equation}
	S_{\mathrm{eff},1} = \frac{\epsilon^p}{2^{p+1} p^2} \, ,
\end{equation}
presented in \cref{eq:Seff1QM} in the main text.

\bibliographystyle{inspire}
\bibliography{refs, refs_noninspire}

\end{document}